\documentclass{article}
\usepackage[T1]{fontenc} %
\usepackage[utf8]{inputenc} %
\usepackage{ismir,amsmath,cite,url}
\usepackage{float}
\usepackage{longtable}
\usepackage{algorithm}
\usepackage{algorithmic}

\usepackage{graphicx}

\usepackage{amsmath}
\usepackage{bbm}
\usepackage{booktabs}
\usepackage{amsfonts}
\usepackage{selectp}
\usepackage{graphicx}
\usepackage{color}
\usepackage{enumitem}
\usepackage{amsfonts}
\usepackage{amssymb}
\usepackage{amsmath}
\usepackage{bm}
\usepackage{lipsum}
\usepackage{graphicx}
\usepackage[dvipsnames]{xcolor}
\usepackage{ragged2e}
\usepackage{multirow}
\usepackage{pifont}

\usepackage{multicol}
\usepackage{caption}
\usepackage{subcaption}
\usepackage{mathtools}
\usepackage{microtype}
\usepackage{changepage}
\usepackage{graphicx}
\newcommand{\eref}[1]{(\ref{#1})}

\newcommand{\sref}[1]{Section~\ref{#1}}
\newcommand{\fref}[1]{Fig.~\ref{#1}}
\newcommand{\tref}[1]{Table~\ref{#1}}

\usepackage{pifont}
\usepackage{amsmath}
\usepackage{amssymb}
\usepackage{mathtools}
\usepackage{amsthm, bm}
\usepackage{color}

\theoremstyle{plain}

\theoremstyle{definition}

\theoremstyle{remark}

\usepackage[colorinlistoftodos,textwidth=3.0cm,textsize=tiny]{todonotes}

\newcommand{\acrolong}{\textbf{D}istilled \textbf{D}iffusion \textbf{I}nference-\textbf{T}ime $\boldsymbol{T}$-\textbf{O}ptimization}

\newcommand{\acro}{DITTO-2}
\usepackage{lineno}

\DeclarePairedDelimiter\floor{\lfloor}{\rfloor}
\title{DITTO-2: Distilled Diffusion Inference-Time T-Optimization\\ for Music Generation}

\multauthor
{Zachary Novack$^1$ \hspace{1cm} Julian McAuley$^1$ \hspace{1cm} Taylor Berg-Kirkpatrick$^1$ \hspace{1cm} Nicholas Bryan$^2$}{\\
 $^1$ University of California -- San Diego\\
$^2$ Adobe Research\\
{\tt\small znovack@ucsd.edu, njb@ieee.org}
}

\begin{document}

\maketitle

\begin{abstract}
Controllable music generation methods are critical for human-centered AI-based music creation, but are currently limited by speed, quality, and control design trade-offs.
Diffusion inference-time T-optimization (DITTO), in particular, offers state-of-the-art results, but is over 10x slower than real-time, limiting practical use. We propose \acrolong{} (or \acro{}), a new method to speed up inference-time optimization-based control and unlock faster-than-real-time generation for a wide-variety of applications such as music inpainting, outpainting, intensity, melody, and musical structure control. Our method works by 
(1) distilling a pre-trained diffusion model for fast sampling via an efficient, modified consistency or consistency trajectory distillation process (2) performing inference-time optimization using our distilled model with one-step sampling as an efficient surrogate optimization task and (3) running a final multi-step sampling generation (decoding) using our estimated noise latents for best-quality, fast, controllable generation. Through thorough evaluation, we find our method not only speeds up generation over 10-20x, but simultaneously improves control adherence and generation quality all at once. Furthermore, we apply our approach to a new application of maximizing text adherence (CLAP score) and show we can convert an unconditional diffusion model without text inputs into a model that yields state-of-the-art text control. Sound examples can be found at \url{https://ditto-music.github.io/ditto2/}.
\end{abstract}

\section{Introduction}\label{sec:introduction}
\begin{figure}[ht!]
    \centering
    \includegraphics[width=0.45\textwidth, trim={0cm .25cm 0 1.0cm},clip]{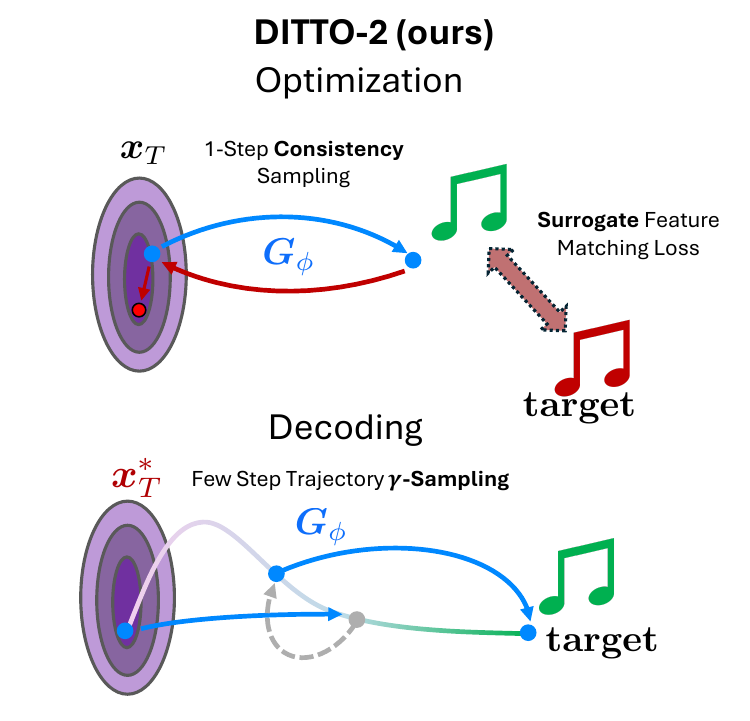}
    \caption{\acro{}: \acrolong{}. We speed up diffusion inference-time optimization-based  music generation by 10-20x while improving control and audio quality. (Top) We use diffusion distillation to speed up performance (optimize with 1-step sampling). (Bottom) We then run multi-step sampling for final higher-quality generation (decoding).
    }
    \label{fig:mainfig}
\end{figure}

Audio-domain text-to-music (TTM) methods~\cite{agostinelli2023musiclm, copet2023simple, forsgren2022riffusion, stableaudio, liu2023audioldm, liu2023audioldm2} have seen rapid development in recent years and show great promise for music creation. Such progress has been made possible through the development of diffusion models~\cite{ho2020denoising, song2020denoising, rombach2022high}, language models~\cite{agostinelli2023musiclm, copet2023simple},  latent representations of audio~\cite{kumar2019melgan, zeghidour2021soundstream, defossez2022highfi, kumar2023high} and text-based control. Such control, however, can be limiting for creative human-centered AI music applications, motivating more diverse and advanced control (e.g.,~melody) that target's fine-grained aspects of musical composition.

Recent control methods that go beyond text-control fall into training-based and training-free methods. Training-based methods like Music-ControlNet~\cite{Wu2023MusicCM} fine-tune DMs with additional adaptor modules that can add time-dependent controls over melody, harmony, and rhythm, offering strong control at the cost of hundreds of GPU hours of fine-tuning for each control. With training-free methods, in particular the class of inference-time \emph{guidance} methods~\cite{Levy2023ControllableMP, yu2023freedom}, the diffusion sampling process is guided at each step using the gradients of a target control $\nabla_{\bm{x}_t} \mathcal{L}(\hat{\bm{x}}_0(\bm{x}_t))$, where $\hat{\bm{x}}_0(\bm{x}_t)$ is a 1-step approximation of the final output. While training-free, the reliance on approximate gradients limits performance~\cite{Novack2024Ditto}. Finally, inference-time optimization (ITO) methods~\cite{Wallace2023EndtoEndDL, karunratanakul2023optimizing} like DITTO~\cite{Novack2024Ditto} offer state-of-the-art (SOTA) control without the need for large-scale fine-tuning via optimizing for noise latents, but suffer from slow inference speeds (10-20x slower than real-time)~\cite{Novack2024Ditto}.

In this work, we propose \acrolong{} (or \acro{}), a new method for speeding up ITO-based methods by over an order of magnitude for faster-than-real-time generation for a wide-variety of controllable generation tasks including inpainting, outpainting, intensity, melody, and musical structure control. Our method works via 1) distilling a pre-trained diffusion model for fast sampling via an efficient, modified consistency model (CM)~\cite{song2023consistency} or consistency trajectory model (CTM)~\cite{Kim2023ConsistencyTM} distillation process (only 32 GPU hours on a 40GB A100), 
(2) performing inference-time optimization using our distilled model with a 1-step \emph{surrogate} objective, and (3) running a final multi-step sampling generation (decoding) using our estimated noise latents for final best-quality results as shown in~\fref{fig:mainfig}.
We find our approach accelerates optimization 10-20x, improves control, and improves audio quality at all once. Furthermore, we apply our approach to maximize text adherence (CLAP score) and show we how  an unconditional diffusion model trained without text inputs can yields SOTA text control. 

\section{Background}\label{sec:relworks}

\subsection{Diffusion-Based Music Generation}
Audio-domain music generation has become tractable through diffusion-based methods, popularized with models such as Riffusion~\cite{forsgren2022riffusion}, MusicLDM~\cite{chen2023musicldm}, and Stable Audio~\cite{stableaudio}. Diffusion Models (DMs)~\cite{song2020denoising, dhariwal2021diffusion} are defined using a closed-form forwards process, where input audio is iteratively noised according to a Gaussian Markov Chain. DMs then learn to approximate the score of the probability distribution of the reverse process $\nabla_{\bm{x}_t}\log q(\bm{x}_t)$ using a noise prediction model $\bm{\epsilon}_\theta$, which progressively denoises a random initial latent $\bm{x}_T \sim \mathcal{N}(0, I)$ to generate new data $\bm{x}_0$. For audio-domain DMs, diffusion is performed over spectrograms~\cite{Wu2023MusicCM} or on the latent representations of an audio-based VAE~\cite{forsgren2022riffusion, chen2023musicldm, stableaudio, Lam2023EfficientNM}, with an external vocoder used to translate spectrograms back to the time domain. Though DMs are efficiently trained by a simple MSE score matching objective~\cite{song2020denoising, Song2020ScoreBasedGM}, sampling from DMs requires running the denoising process for 100s of iterations (calls to $\bm{\epsilon}_\theta$), and have slower inference than VAEs or GANs~\cite{Pasini2022MusikaFI}.

\subsection{Fast Diffusion Sampling}
Fast diffusion sampling is critical. DDIM~\cite{song2020denoising} or DPM-Solver~\cite{Lu2022DPMSolverFS} accelerate DMs to sample in only 10-50 sampling steps.  
To truly increase speed, however, \emph{distillation} can be used to produce a model that can sample in a \emph{single} step~\cite{song2023consistency, luo2023latent, Sauer2023AdversarialDD, Kim2023ConsistencyTM}. Two promising DM distillation methods include consistency models (CM)~\cite{song2023consistency} and consistency trajectory models~\cite{Kim2023ConsistencyTM}. The goal of CMs is to distill a base DM $\bm\epsilon_\theta$ into a new 1-step network $\bm{x}_0 = \bm{G}_\phi(\bm{x}_t, \bm{c})$ that satisfies the consistency property $\forall t, t' \in [T, 0], \bm{G}_\phi(\bm{x}_t, \bm{c}) = \bm{G}_\phi(\bm{x}_{t'}, \bm{c})$ or that every point along the diffusion trajectory maps to the same output. Formally, CMs are distilled by enforcing local consistency between the learnable $\bm{G}_\phi$ and an exponential moving average (EMA) copy $\bm{G}_{\phi^-}$:
\begin{equation}\label{eq:cm}
    \mathbb{E}_{t \sim T, (\bm{x}, \bm{c}) \sim \mathcal{D}}\| \bm{G}_\phi(\bm{x}_t, \bm{c}) - \bm{G}_{\phi^-}(\Theta(\bm\epsilon_\theta, \bm{x}_t, \bm{c}) , \bm{c})\|_2^2,
\end{equation}
where $\Theta(\bm\epsilon_\theta, \bm{x}_t, \bm{c})$ denotes one sampling step from $\bm{x}_t$ to $\bm{x}_{t-1}$ using the \emph{frozen} teacher model $\bm\epsilon_\theta$ and some sampling algorithm (e.g.~DDIM).

\begin{figure}
    \centering
    \includegraphics[width=0.4\textwidth]{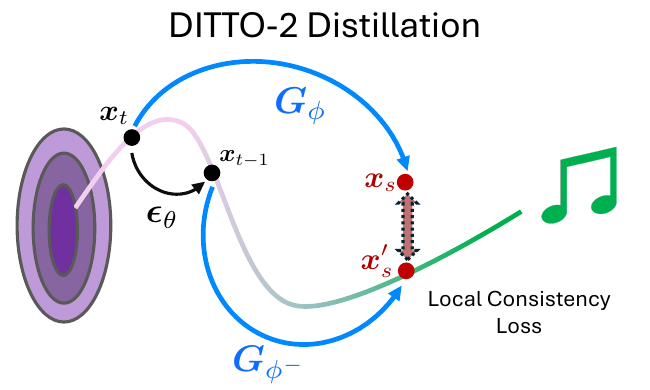}
    \caption{CTM Distillation for \acro{}. We distill $\bm{G}_\phi$ by minimizing the distance between the jump from $\bm{x}_t$ to $\bm{x}_s$ and $\bm{x}_{t-1}$ to $\bm{x}_s$, where $\bm{x}_{t-1}$ is generated by sampling with the base model $\bm\epsilon_\theta$.}
    \label{fig:distill}
\end{figure}

CMs are not perfect, however, and one-step performance lags behind DM quality~\cite{luo2023latent}. Multi-step ``ping-pong'' sampling also does not reliably increase quality due to compound approx. errors in each renoising step.
CTMs~\cite{Kim2023ConsistencyTM}, on the other hand, are designed to fix this problem. CTMs bridge the gap between CMs and DMs by distilling a model $\bm{x}_s = \bm{G}_\phi(\bm{x}_t, \bm{c}, w, t, s)$ that can jump from \emph{anywhere to anywhere} along the diffusion trajectory as shown in~\fref{fig:distill}. CTMs then use $\gamma$-sampling to interpolate between few-step deterministic sampling along the trajectory ($\gamma = 0$) and CM's ``ping-pong'' sampling ($\gamma = 1$), allowing a way to balance sampling stochasticity with overall quality. To our knowledge, CTM distillation is unexplored for audio, and  CM distillation has only been applied to general audio~\cite{bai2023accelerating}. 

\subsection{Diffusion Inference-time Optimization}
\begin{figure}[t!]
    \centering
    \includegraphics[width=0.45\textwidth, trim={0cm 13cm 0 1cm},clip]{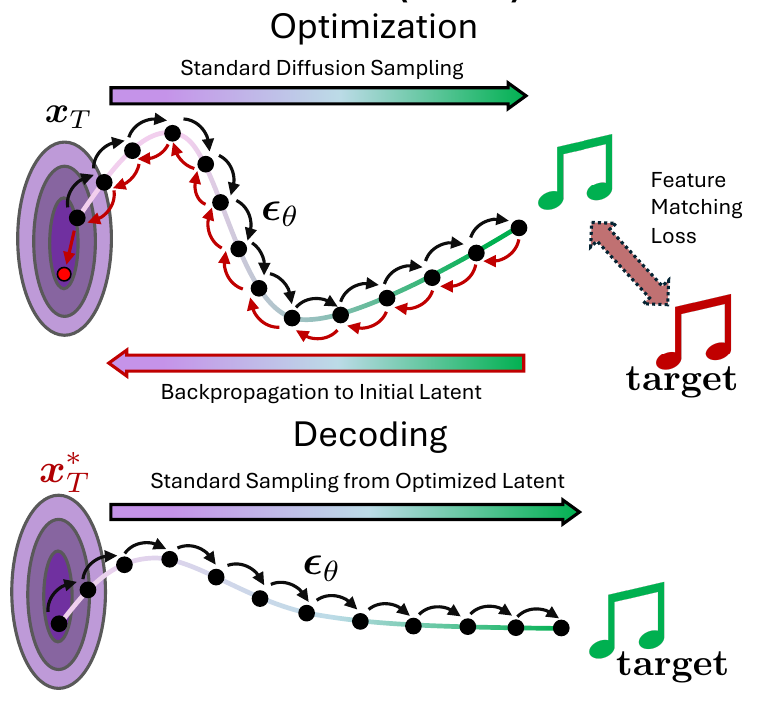}
    \vspace{-.5cm}
    \caption{(Top) Baseline DITTO runs optimization over a multi-step sampling process to find an initial noise latent to achieve a desired stylized output, incurring a large speed cost. (Bottom) When generating the final output (decoding), the same multi-step diffusion sampling process is used.}
    \label{fig:mainfig2}
\end{figure}
Diffusion inference-time optimization (DITTO)~\cite{Wallace2023EndtoEndDL, karunratanakul2023optimizing, Novack2024Ditto} is a general-purpose framework to control diffusion models at inference-time. The work is based on the observation that the \emph{initial noise latent} $\bm{x}_T$, traditionally thought of as a random seed, encodes a large proportion of the semantic content in the generation outputs~\cite{Si2023FreeUFL, Novack2024Ditto}. 
Thus, we can search for an initial noise latent of the diffusion generation process via optimization to achieve a desired stylized output as shown in~\fref{fig:mainfig2}. We do this by defining a differentiable feature extraction function (e.g.~chroma-based melody extraction) $f(\cdot)$, a matching loss function $\mathcal{L}$ (e.g.~cross entropy), a target feature $\bm{y}$, and optimize $\bm{x}_T$:
\begin{align}
    \boldsymbol{x}_T^* &= \arg \min_{\boldsymbol{x}_T} \mathcal{L}\left(f(\boldsymbol{x}_0), \boldsymbol{y}\right) \label{eq:problem}\\
    \bm{x}_0 &= \Theta_T(\bm{\epsilon}_\theta, \bm{x}_T, \bm{c}),\label{eq:problem2}
\end{align}
where $\Theta_T(\bm{\epsilon}_\theta, \bm{x}_T, \bm{c})$ denotes $T$ calls of the model using any sampler $\Theta$. In practice, DITTO is run with a fixed budget of $K$ optimization steps using a standard optimizer (i.e.~Adam). This approach allows for any control that can be parameterized differentiably, including melody, intensity, and musical structure, as well as editing tasks like inpainting and outpainting. For brevity, we combine Eq.~\ref{eq:problem} and Eq.~\ref{eq:problem2} into the shorthand $\boldsymbol{x}_T^* = \arg \min_{\boldsymbol{x}_T} \mathcal{L}_\theta^{(T)}\left(\bm{x}_T\right)$.

The downside of DITTO, however, is that it is slow. We need to backpropagate through the \emph{entire} sampling process for each of the $K$ optimization steps and use memory management techniques like gradient checkpointing~\cite{chen2016training} or invertible networks~\cite{Wallace2023EndtoEndDL} to handle large memory use that slows down generation. 
The overall cost of running a single ITO generation is on the order of $4KT$: $T$-step diffusion chain for $K$ opt. steps, with a factor of 2 from gradient management and 2 from using classifier-free guidance (CFG)~\cite{ho2022classifier} to improve quality.

\section{Method}\label{sec:method}

\subsection{Overview}
We seek to dramatically speed up the diffusion ITO process to achieve controllable music generation for near-interactive rate music co-creation.  To do so, we focus on three critical methodological improvements. First, we leverage \textbf{diffusion distillation} to speed up the diffusion sampling with an efficient, modified distillation process designed to be used together with ITO methods. Second, we introduce \textbf{\emph{surrogate} optimization}, or the idea of decoupling the task of estimating noise latents from the task of rendering a final output or decoding, which allows us to leverage both fast sampling for optimization for control estimation and multi-step sampling for final, high-quality generation. Third, we combine diffusion distillation with surrogate optimization within the DITTO framework and produce a new, more efficient diffusion inference-optimization algorithm (no gradient checkpointing) as found in~\sref{sec:algo}. 

\subsection{Acceleration through Diffusion Distillation}
The clearest way to speed up ITO is to simply reduce the number of diffusion sampling steps $T$. From initial experiments, however, we found that (1) reducing the number of sampling steps $T$ degrades overall generation quality~\cite{song2020denoising}, (2)  quality degradation makes the optimization gradients weaker (as the outputs are less semantically coherent), leading to control degradation, and (3) achieving close to real-time performance requires $<4$ sampling steps, which produce fully incoherent results on standard DMs. Thus, we employ distillation to speed up the diffusion process~\cite{luo2023latent}.

First, we develop CM distillation~\cite{song2023consistency, bai2023accelerating, luo2023latent} for ITO-based controllable music generation. For CM distillation, we follow past work~\cite{luo2023latent} in training recipe, optimizing~\eref{eq:cm}, and 
also learn an explicit embedding for the CFG scale $w$ in the model $\bm{G}_\phi(\bm{x}_t, \bm{c}, w)$ during distillation following \cite{luo2023latent}. By distilling CFG, we are able to half the number of total model calls per distilled diffusion sampling step. Once distilled, $\bm{G}_\phi$ jumps from $\bm{x}_T$ to $\bm{x}_0$, allowing for deterministic 1-step sampling and stochastic multistep-sampling by repeatedly \emph{renoising} with some $\epsilon \sim N(0, I)$ back to $\bm{x}_{t-1}$.  

Second, we develop CTM distillation~\cite{Kim2023ConsistencyTM} for ITO-based controllable music generation. CTM distillation offers more advantageous speed vs. quality design trade-offs, but comes at a cost of a more complex training procedure. In more detail, CTM distillation normally involves an expensive soft-consistency loss in the data domain with added GAN and score-matching loss terms.  
As we aim to distill our base model for \emph{surrogate} optimization (see Sec.~\ref{sec:surr}), we are able to simplify and speed up the CTM distillation process. 
First, we remove the image-domain GAN loss to reduce complexity of developing an audio-based GAN loss. Second, we use the consistency term from CTM in local-consistency form~\cite{zheng2024trajectory}:
\begin{align}\label{eq:ctm}
    \mathbb{E}_{t, s \sim T, (\bm{x}, \bm{c}) \sim \mathcal{D}}&\| \bm{G}_\phi(\bm{x}_t, \bm{c}, w, t, s) - \nonumber\\ &\bm{G}_{\phi^-}(\Theta(\bm\epsilon_\theta, \bm{x}_t, \bm{c}) , \bm{c}, w, t-1, s)\|_2^2
\end{align}
Third, we use the 1-step Euler parameterization of $\bm{G}_\phi$ from \cite{Kim2023ConsistencyTM}'s Appendix, which avoids explicitly learning additional parameters for the target step $s$ in order to accelerate distillation. Finally, we upgrade CTM's unconditional framing for conditional diffusion by incorporating $\bm{c}$ into the distillation procedure, and adding $w$ directly into the model to distill the CFG weight into an explicit parameter following past work~\cite{luo2023latent}, resulting in CFG control at inference but without double the complexity. In total, we perform distillation in as few as 32 GPU hours on an A100, the fastest trajectory-based distillation to our knowledge~\cite{Kim2023ConsistencyTM, zheng2024trajectory}.

\subsection{Surrogate Optimization}\label{sec:surr}
\begin{algorithm}[t!]
    \caption{ 
    \acrolong{} (\acro{})}
    \label{alg:aditto}
    \begin{algorithmic}[1]
    \INPUT: $\bm{G}_\phi$, feature extractor $f$, loss $\mathcal{L}$, target $\bm{y}$, starting latent $\bm{x}_T$, text $\bm{c}$, optimization steps $K$, optimizer $g$, decoding steps $\{\tau_0, \dots, \tau_M\}$, $\gamma$, CFG weight $w$ 
        \STATE // Optimization Loop
        \FOR{$K$ iterations}
            \STATE $\boldsymbol{x}_0 = \bm{G}_\phi(\boldsymbol{x}_T, \bm{c}, w, T, 0)$ %
            \STATE $\hat{\boldsymbol{y}} = f(\boldsymbol{x}_0)$ %

            \STATE $\boldsymbol{x}_T \leftarrow \boldsymbol{x}_T - g(\nabla_{\boldsymbol{x}_T}\mathcal{L}(\hat{\boldsymbol{y}}, \boldsymbol{y}))$ %
        \ENDFOR
        \STATE // Decoding Loop
        \STATE $\bm{x}_{t} \leftarrow \bm{x}_T$
        \FOR{$t=M$ to 1}
            \STATE $\hat\tau_{t-1} = \sqrt{1 - \gamma^2}\tau_{t-1}$ 
            \STATE $\bm{x}_{t-1} = \bm{G}_\phi(\bm{x}_{t-1}, \bm{c}, w, \tau_{t}, \hat\tau_{t-1}) + \gamma \tau_{t-1} \epsilon$ 
        \ENDFOR

    \OUTPUT: $\boldsymbol{x}_0$
    
\end{algorithmic}
\end{algorithm}

Given a distilled CM or CTM model, we seek to minimize our inference runtime and maximize control adherence and audio quality. The obvious choice to minimize runtime is to use our distilled model with one-step sampling, but this results in limited audio quality and text-control. To solve this, we first split the ITO process into two separate phases: \textbf{optimization}, i.e.~the nested loop of optimizing the initial latent over $M$-step multistep sampling, and \textbf{decoding}, i.e.~the final $T$-step sampling process from the optimized latent $\bm{x}_T^*$, where $M=T$ in all past work \cite{Novack2024Ditto, Wallace2023EndtoEndDL}.
In this light, it is clear that the optimization phase is mostly responsible for the  control strength and runtime, while decoding is generally responsible for final output quality. 

Thus, we fix our final decoding process as multi-step sampling with $T$ steps. Then, we perform control optimization over a \textbf{surrogate} objective  $\hat{\bm{x}}_T^* = \arg \min_{\bm{x}_T}\mathcal{L}_\phi^{(M)}\left(\bm{x}_T\right)$ using some model $\bm\epsilon_\phi$ and $M \ll T$, where our surrogate is more efficient but yields approx. equal latents to our original objective
\begin{align}
    \arg \min_{\bm{x}_T }\mathcal{L}_\phi^{(M)}\left(\bm{x}_T\right) &\approx \arg \min_{\bm{x}_T } \mathcal{L}_\theta^{(T)}\left(\bm{x}_T\right).
\end{align}
A natural candidate for a surrogate model would be the base DM $\bm\epsilon_\theta$ with fewer sampling steps. DM performance, however, becomes fully incoherent as $M \xrightarrow{} 1$~\cite{song2023consistency, Kim2023ConsistencyTM}, causing a significant  domain-gap when $M < T$. Alternatively, our distilled models are naturally strong surrogates:
\begin{itemize}
    \item One-step outputs are generally coherent unlike in base DMs, resulting in more stable gradients when $M=1$.
    \item Since distilled models excel at few-step sampling (i.e. $<8$), the control domain gap between $M$ and $T$ can be reduced while still ensuring coherent outputs.
    \item CTMs can increase quality with more sampling.
\end{itemize}
As a result, we use a CM or CTM $\bm{G}_\phi$ as our surrogate, optimize with $M=1$, and decode with $T \in [1,8]$.

\subsection{Complete Algorithm}\label{sec:algo}
Given our efficient CTM-based distillation process, and our surrogate objective, we propose a new ITO algorithm for controllable music generation in Alg.~\ref{alg:aditto}. Here, we run optimization to estimate control parameters using our surrogate 1-step objective. Then, we use the optimized latent $x_T^*$ and decode from our surrogate model with $T$ steps using either multi-step CM Sampling (i.e.~$\gamma = 1$) or CTM $\gamma$-sampling ($\gamma < 1$). Beyond decoupling optimization and decoding, we also eliminate any use gradient checkpointing found in the original DITTO method~\cite{Novack2024Ditto}. In total, we reduce the ITO speed from $4KT$ costly operations for DITTO to $K + T$.

\section{Experiments}
To evaluate our proposed method, we follow the evaluation protocol used for DITTO~\cite{Novack2024Ditto} for intensity, melody, music structure, inpainting, and outpainting as described below. Before the full breadth of application tests, however, we explore our design space by comparing different distillation techniques and surrogate options on the task of intensity control. We further conclude with an experiment showing an adaptive sampling surrogate scheme as well a new experiment on maximizing text-adherence (CLAP score).

\subsection{Controllable Generation Evaluation Protocol}
We benchmark our method on five controllable music generation tasks from DITTO~\cite{Novack2024Ditto} including: 
\begin{itemize}
    \item \textbf{Intensity Control} \cite{Wu2023MusicCM, Novack2024Ditto}: Here, we control the time-varying volume and overall semantic density to some target intensity curve $\bm{y}$ using the extractor $f(\bm{x}_0) \coloneqq \bm{w} \ast 20\log_{10}(\texttt{RMS}(\mathbf{V}(\bm{x}_0)))$ (i.e.~the smoothed RMS energy of the vocoded model outputs) and $\mathcal{L} = ||f(\bm{x}_0)- \bm{y}||_2^2$.
    \item \textbf{Melody Control} \cite{copet2023simple, Wu2023MusicCM, Novack2024Ditto}: We control the model outputs to match a given target melody $\bm{y} \in \{1, \dots, 12\}^{N \times 1}$ (where $N$ is number of frames) using the chromagram of the model outputs $f(\bm{x}_0) = \log(\mathbf{C}(\mathbf{V}(\bm{x}_0)))$ and $\mathcal{L} = \text{NLLLoss}(f(\bm{x}_0), \bm{y})$.
    \item \textbf{Musical Structure Control} \cite{Novack2024Ditto}: We control the overall timbral structure of the model outputs by regressing the Mel Frequency Cepstrum Coefficient (MFCC) self-similarity (SS) matrix $f(\bm{x}_0) = \mathbf{T}(\bm{x}_0) \mathbf{T}(\bm{x}_0)^\top$ against a target SS matrix $\bm{y}$ like ``ABA" form with $\mathcal{L} = ||f(\bm{x}_0) - \bm{y}||_2^2$.
    \item \textbf{Inpainting and Outpainting} \cite{Levy2023ControllableMP, Novack2024Ditto}: Given music $\bm{x}_{\text{ref}}$, we can continue (outpainting) or infill (inpainting) $\bm{x}_{\text{ref}}$ by matching the model outputs over $o$-length overlap regions $f(\bm{x}_0)\coloneqq \mathbf{M}_{\text{gen}} \odot \bm{x}_0$ to the reference $\bm{y}= \mathbf{M}_{\text{ref}} \odot \bm{x}_{\text{ref}}$ and $\mathcal{L} = ||f(\bm{x}_0)- \bm{y}||_2^2$.
\end{itemize}
For brevity, we focus on the $o=1$ case for outpainting and inpainting (i.e.~a gap of 4 seconds) and omit looping given its equivalence. See \cite{Novack2024Ditto} for a more thorough description. 

\subsection{Pre-training and Distillation Details}
For our base DM, we follow a similar setup and model design to DITTO~\cite{Novack2024Ditto}. Specifically, we train a 41M parameter Stable Diffusion-style 2D UNet directly over 6-second mel-spectrograms trained on $\approx 1800$ hours of licensed music, and MusicHiFi~\cite{Zhu2024MusicHiFiFH} as the vocoder. The base model is trained with genre, mood, and tempo tags similar to \cite{dhariwal2020jukebox} rather than full text descriptions. Both the CM and CTM surrogate models are distilled using a maximum of $T=20$ sampling steps, evenly spaced across the trajectory for 4 hours across 8 A100 40GB GPUs on the same data. For \acro{}, we use Adam. During CTM $\gamma$-sampling, we set $\gamma \in [0.05, 0.35]$ as empirically we found that using deterministic $\gamma = 0$ resulted in noticable audio artifacts that degrade overall quality.

\subsection{Metrics}
For all tasks, we report the Frechet Audio Distance (FAD) with the CLAP~\cite{wu2023large} music backbone (as the standard VGGish backbone poorly correlates with human perception of quality~\cite{fadtk}) and the CLAP Score, which measure overall audio quality and text relevance respectively across 2.5K generations. FAD is calculated with MusicCaps as the reference ~\cite{agostinelli2023musiclm} dataset. Since our base model uses tags rather than captions, we convert each tag set into captions for CLAP Score calculation using the format \emph{``A [mood] [genre] song at [tempo] beats per minute.''} Additionally, we report the MSE to the control target for intensity and structure control, and the overall accuracy for melody control.

\section{Results}

\subsection{Design Exploration Results}
\begin{figure}[t!]
    \centering
    \hspace*{-.4cm}    
    \includegraphics[width=0.5\textwidth, trim={0.1cm 0 0 0},clip]{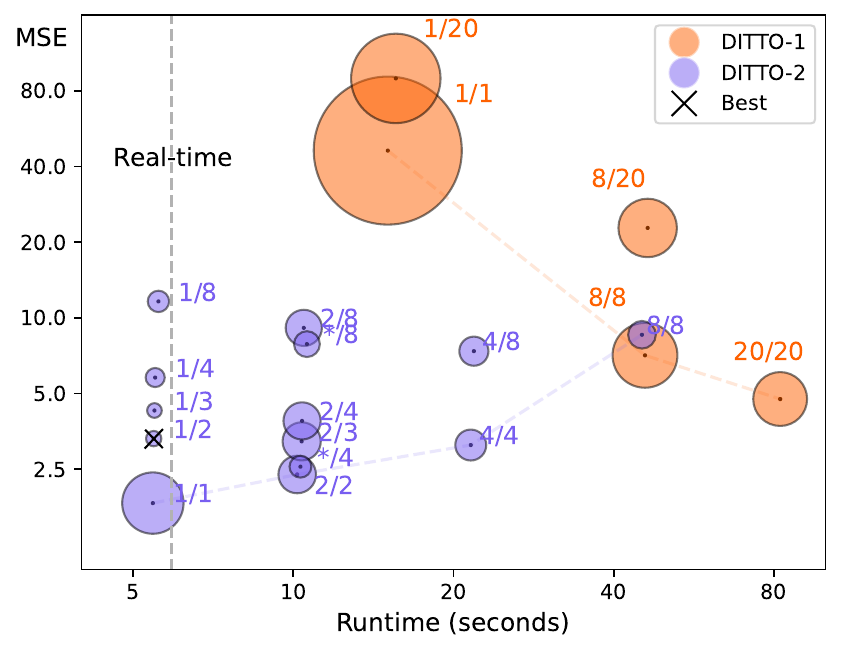}
    \caption{\textbf{\acro{} inference speed vs. control MSE vs. audio quality (FAD, denoted by size, smaller is better)}. 
    Dashed line denotes the cutoff for real-time performance, color denotes ITO method, and subscripts denote number of sampling steps during optimization / final decoding. Applied to intensity control. Trends also hold for CLAP score.}
    \label{fig:headline}
\end{figure}
\begin{figure}[t!]
    \centering
    \includegraphics[width=0.48\textwidth, trim={0.1cm 0 0 0},clip]{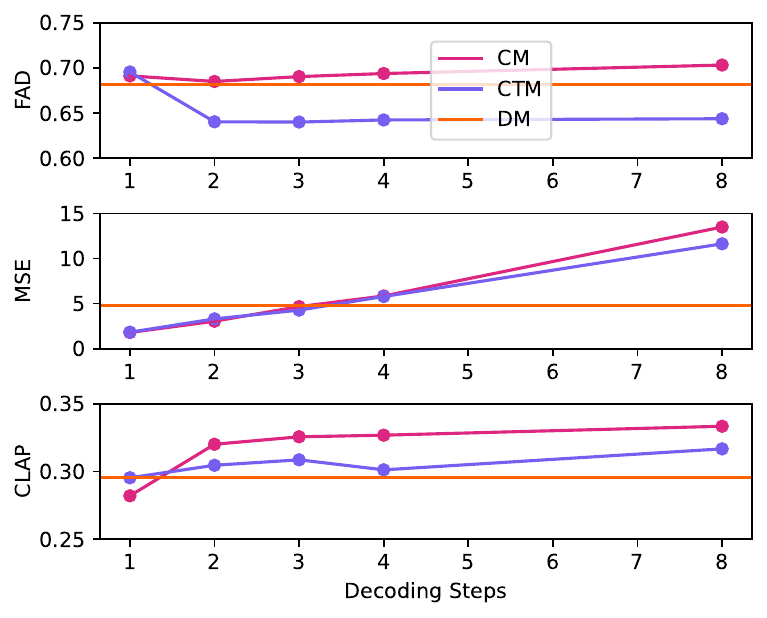}
    \caption{FAD, MSE, and CLAP results on Intensity Control for 1-step optimization, where orange lines denote baseline 20-step performance. MSE increases with more decoding steps for both CM/CTM given the domain gap though beats the baseline with $<4$ steps. CM is unable to beat baseline quality due to accumulated errors in multi-step sampling, while multi-step CTM achieves SOTA quality.}
    \label{fig:main_res}
\end{figure}

We study our design space for ITO via a case study on the task of intensity control. Notably, we compare DITTO with our proposed approach using CM and CTM distilled models in~\fref{fig:headline}.  We show runtime in seconds (x-axis), control MSE (y-axis), and FAD (point size) for an array of $(M, T)$ combinations for our base DM, as well as our distilled CTM models (i.e. DM-8/20 corresponds to the base DM with $M=8, T=20$). We find that our distilled models are over 10x faster than the standard DITTO (20, 20) configuration, while simultaneously achieving both better audio quality and control adherence. To understand these trends more in depth, specifically as we vary the number of decoding steps for our distilled models, we show both FAD (top), MSE (middle), and CLAP (bottom) in~\fref{fig:main_res} as a function on number of decoding steps with $M=1$, where the orange line denotes the baseline DITTO results with $M=T=20$. 

A few key points of \acro{} are visible here. Notably, both CM and CTM distilled models are able to achieve \emph{better} control adherence than the base performance, as the shorter optimization process allows convergence to happen more effectively. Additionally, we find that CTM is clearly stronger than CM in terms of quality, as CTM is able to cleanly trade-off quality for control adherence (as sampling with more steps with $M=1$ introduces a domain gap) and even \emph{improve} baseline quality, while CM exhibits no real trend quality when sampling more due to its accumulated errors in multi-step sampling. In particular, CTM with $M=1, T=2$ achieves SOTA control adherence and FAD with faster than real-time. CM and CTM multi-step sampling also improves text relevance above the base DM.

\subsection{Benchmark Results}
We show full results on our suite of controllable music generation benchmarking results in~\tref{tab:allres}. Here, we compare baseline DITTO with \acro{}, where we display results for the best performing $(M, T)$ setup, in terms of audio quality and control adherence jointly, for each experiment for CM and CTM.  As a whole, \acro{} achieves comparable or better performance than DITTO on all tasks with an 10-20x speedup, clearing the way for near real-time inference-time controllable music generation.

\begin{table}[t!]
    \centering
    \begin{tabular}{l|cccc}
    \toprule
         \multicolumn{5}{c}{\textbf{Intensity}}\\
         Method&  Time (s)&  FAD&  CLAP& MSE\\
         \midrule
         DITTO&  82.192&  \underline{0.682}&  0.296& 4.758\\
         \acro{} (CM)&  \textbf{5.206}&  0.685&  \textbf{0.320}& \textbf{3.055}\\
         \acro{} (CTM)&  \underline{5.467}&  \textbf{0.640}&  \underline{0.309}& \underline{3.311}\\
         \midrule
         \multicolumn{5}{c}{\textbf{Melody}}\\
         Method&  Time (s)&  FAD&  CLAP& Acc.\\
         
         \midrule
         DITTO&  230.780&  0.699&  \underline{0.283}& \underline{82.625}\\
         \acro{} (CM)&  \textbf{21.867}&  \textbf{0.697}&  \textbf{0.303}& 81.577\\
 \acro{} (CTM)& \underline{22.501}& \underline{0.698}& 0.273&\textbf{85.226}\\
 \midrule
 \multicolumn{5}{c}{\textbf{Musical Structure}}\\
 Method& Time (s)& FAD& CLAP&MSE\\
 \midrule
 DITTO& 245.295& \textbf{0.632}& \textbf{0.281}&0.024\\
 \acro{} (CM)& \textbf{11.381}& 0.669& \underline{0.234}&\textbf{0.020}\\
 \acro{} (CTM)& \underline{11.749}& \underline{0.658}& 0.226&\underline{0.022}\\
 \midrule
 \multicolumn{5}{c}{\textbf{Outpainting}}\\
 Method& Time (s)& FAD& CLAP&\\
  \midrule
 DITTO& 144.437& 0.716& \underline{0.343}&\\
 \acro{} (CM)& \textbf{6.658}& \underline{0.694}& 0.319&\\
 \acro{} (CTM)& \underline{7.098}& \textbf{0.680}& \textbf{0.347}&\\
  \midrule
 \multicolumn{5}{c}{\textbf{Inpainting}}\\
 Method& Time (s)& FAD& CLAP&\\
  \midrule
 DITTO& 145.486& 0.690& \underline{0.339}&\\
 \acro{} (CM)& \textbf{6.744}& \underline{0.689}& \textbf{0.358}&\\
 \acro{} (CTM)& \underline{6.814}& \textbf{0.660}& 0.337&\\
 \bottomrule
    \end{tabular}
    \caption{Controllable generation benchmark results. Best performing configuration for each \acro{} setup across five unique tasks. Both CM and CTM results yield excellent results with 10-20x speed ups.}
    \label{tab:allres}
\end{table}

\subsection{Variable Compute Budget Optimization}

\begin{table}[t!]
    \centering
    
    \begin{tabular}{cc|cccc}
    \toprule
         $M$&  $T$&  Runtime&  FAD&  CLAP& MSE\\
         \midrule
         1& 1&  5.447&  0.696&  0.295& 1.835\\
 1& 2& 5.467& 0.640& 0.307&3.311\\
 1& 4& 5.502& 0.643& 0.301&5.792\\
         2&  2&  10.171&  0.659&  0.281& 2.384\\
         2&  4&  10.387&  0.658&  0.296& 3.894\\
         Adaptive &  4&  10.315&  0.644&  0.296& 2.561\\
         \bottomrule
    \end{tabular}
    \caption{Intensity control results with various $(M,T)$ options including an adaptive sampling during optimization.}
    \label{tab:varcomp}
\end{table}

Though we are primarily interested in real-time performance (i.e.~as fast as possible), we additionally investigated how we can use a varying the compute budget during optimization (in terms of runtime). As simply increasing $M$ predictably increases runtime by a multiplicative factor, we designed an \emph{adaptive} schedule for $M$ (denoted as $*$ in Fig.~\ref{fig:headline}) in order to improve downstream decoding performance without increasing runtime significantly. Formally, for $K$ optimization steps, we set the adaptive budget as using $M=1$ for $\floor{\frac{K}{2}}$ iterations, then $M=2$ for $\floor{\frac{3K}{8}}$, and finally $M=4$ for $\floor{\frac{K}{8}}$ iterations, thus allowing a coarse-to-fine optimization process. In~\tref{tab:varcomp}, we can see that using the adaptive schedule only incurs the runtime of the $M=2$ case, yet achieves much better FAD at similar control adherence. This shows that given a more flexible compute budget, using an adaptive $M$ schedule balances~downstream performance better than simply modifying a fixed $M$.

\subsection{Inference-time Optimization of Text-Control}
Past ITO methods for music generation use simple feature extractors $f(\cdot)$ (i.e. chroma or RMS energy)~\cite{Novack2024Ditto} to minimize runtime speed.
Given our new method is much faster, however, we can introduce new bespoke control applications with neural network-based feature extractors. Thus, we propose the task of inference-time \textbf{text similarity} control. We extract the normalized CLAP audio embedding~\cite{wu2023large} of our model outputs $f(\cdot) = \text{CLAP}(\bm{x}_0)$ and, given some natural language caption $\bm{y}$, calculate the cosine distance between the output and the normalized CLAP text embedding of the caption $\mathcal{L}(\bm{x}_0) = 1 -  f(\bm{x}_0)^\top f(\bm{y})$. 

Using FAD and CLAP score as metrics, we benchmark several configurations including our base DM model with tag inputs, \acro{} method with tag inputs, \acro{} method with null tag inputs, MusicGen w/melody (1.5B)~\cite{copet2023simple}, and MusicGen w.o./melody (3.5B)~\cite{copet2023simple}. For models that take input text, we use captions from MusicCaps~\cite{agostinelli2023musiclm} as input and  for models with tag inputs, we convert MusicCaps captions to tags via GPT-4 as done in past work~\cite{Wu2023MusicCM} with tempo extracted from  audio. Furthermore, to ablate whether any part of the tag-conditioned training process influences downstream \acro{} CLAP control, we retrain and distill our base model \emph{without any text input}, which we denoted U-\acro{}.  In~\tref{tab:clap}, we see that \acro{} enables SOTA text relevance compared to MusicGen by an over 54\% relative improvement (large), and notably enables fully-unconditional models to have text control \emph{with no paired music-text training}.

\begin{table}[t!]
    \centering
    \begin{tabular}{lc|cc}
    \toprule
         Method &Condition &  FAD& CLAP \\
         \midrule
         Base TTM &Tags &  0.488 & 0.167 \\
         \acro{} &Tags &  0.456&  0.317\\
 \acro{} & N/A & 0.440&\underline{0.341}\\
 U-\acro{}& N/A & \textbf{0.430}&\textbf{0.347}\\
 MusicGen (1.5B)& Caption & 0.444 & 0.237\\
 MusicGen (3.3B) & Caption & \underline{0.437}& 0.226\\
 \bottomrule
    \end{tabular}
    \caption{Text similarity results. We use \acro{} to maximize CLAP similarity using a fully unconditional pre-trained diffusion model and yield a 54\% relative improvement over past SOTA CLAP score (MusicGen).}
    \label{tab:clap}
\end{table}

\section{Conclusion}
We present \acro{}: \acrolong{}, a new efficient method for accelerating inference-time optimization for fast controllable music generation. By utilizing a modified consistency or consistency trajectory distillation process and performing inference-time optimization on efficient  surrogate objectives, we speed up past ITO methods by over 10-20x while simultaneously improving audio quality and text control. Furthermore, we find we can leverage the efficiency of our new method on new, more complex tasks like text-adherence and show we can convert a fully unconditional diffusion model into a TTM model that yields SOTA results.

\bibliography{ISMIRtemplate}

\begin{thebibliography}{10}
\providecommand{\url}[1]{#1}
\csname url@samestyle\endcsname
\providecommand{\newblock}{\relax}
\providecommand{\bibinfo}[2]{#2}
\providecommand{\BIBentrySTDinterwordspacing}{\spaceskip=0pt\relax}
\providecommand{\BIBentryALTinterwordstretchfactor}{4}
\providecommand{\BIBentryALTinterwordspacing}{\spaceskip=\fontdimen2\font plus
\BIBentryALTinterwordstretchfactor\fontdimen3\font minus \fontdimen4\font\relax}
\providecommand{\BIBforeignlanguage}[2]{{%
\expandafter\ifx\csname l@#1\endcsname\relax
\typeout{** WARNING: IEEEtran.bst: No hyphenation pattern has been}%
\typeout{** loaded for the language `#1'. Using the pattern for}%
\typeout{** the default language instead.}%
\else
\language=\csname l@#1\endcsname
\fi
#2}}
\providecommand{\BIBdecl}{\relax}
\BIBdecl

\bibitem{agostinelli2023musiclm}
A.~Agostinelli, T.~I. Denk, Z.~Borsos, J.~Engel, M.~Verzetti, A.~Caillon, Q.~Huang, A.~Jansen, A.~Roberts, M.~Tagliasacchi \emph{et~al.}, ``Music{LM}: Generating music from text,'' \emph{arXiv:2301.11325}, 2023.

\bibitem{copet2023simple}
J.~Copet, F.~Kreuk, I.~Gat, T.~Remez, D.~Kant, G.~Synnaeve, Y.~Adi, and A.~D{\'e}fossez, ``Simple and controllable music generation,'' in \emph{Neural Information Processing Systems (NeurIPS)}, 2023.

\bibitem{forsgren2022riffusion}
\BIBentryALTinterwordspacing
S.~Forsgren and H.~Martiros, ``{Riffusion: Stable diffusion for real-time music generation},'' 2022. [Online]. Available: \url{https://riffusion.com/about}
\BIBentrySTDinterwordspacing

\bibitem{stableaudio}
\BIBentryALTinterwordspacing
StabilityAI, ``{Stable Audio}: Fast timing-conditioned latent audio diffusion,'' Nov 2023. [Online]. Available: \url{https://stability.ai/research/stable-audio-efficient-timing-latent-diffusion}
\BIBentrySTDinterwordspacing

\bibitem{liu2023audioldm}
H.~Liu, Z.~Chen, Y.~Yuan, X.~Mei, X.~Liu, D.~Mandic, W.~Wang, and M.~D. Plumbley, ``Audio{LDM}: Text-to-audio generation with latent diffusion models,'' in \emph{International Conference on Machine Learning (ICML)}, 2023.

\bibitem{liu2023audioldm2}
H.~Liu, Q.~Tian, Y.~Yuan, X.~Liu, X.~Mei, Q.~Kong, Y.~Wang, W.~Wang, Y.~Wang, and M.~D. Plumbley, ``{AudioLDM} 2: Learning holistic audio generation with self-supervised pretraining,'' \emph{arXiv preprint arXiv:2308.05734}, 2023.

\bibitem{ho2020denoising}
J.~Ho, A.~Jain, and P.~Abbeel, ``Denoising diffusion probabilistic models,'' \emph{Neural Information Processing Systems (NeurIPS)}, 2020.

\bibitem{song2020denoising}
J.~Song, C.~Meng, and S.~Ermon, ``Denoising diffusion implicit models,'' in \emph{International Conference on Learning Representations (ICLR)}, 2020.

\bibitem{rombach2022high}
R.~Rombach, A.~Blattmann, D.~Lorenz, P.~Esser, and B.~Ommer, ``High-resolution image synthesis with latent diffusion models,'' in \emph{IEEE Conference on Computer Vision and Pattern Recognition (CVPR)}, 2022.

\bibitem{kumar2019melgan}
K.~Kumar, R.~Kumar, T.~De~Boissiere, L.~Gestin, W.~Z. Teoh, J.~Sotelo, A.~De~Brebisson, Y.~Bengio, and A.~C. Courville, ``{MelGAN}: Generative adversarial networks for conditional waveform synthesis,'' \emph{Neural Information Processing Systems (NeurIPS)}, 2019.

\bibitem{zeghidour2021soundstream}
N.~Zeghidour, A.~Luebs, A.~Omran, J.~Skoglund, and M.~Tagliasacchi, ``{SoundStream}: An end-to-end neural audio codec,'' \emph{IEEE/ACM Transactions on Audio, Speech, and Language Processing (TASLP)}, 2021.

\bibitem{defossez2022highfi}
A.~Défossez, J.~Copet, G.~Synnaeve, and Y.~Adi, ``High fidelity neural audio compression,'' \emph{arXiv preprint arXiv:2210.13438}, 2022.

\bibitem{kumar2023high}
R.~Kumar, P.~Seetharaman, A.~Luebs, I.~Kumar, and K.~Kumar, ``High-fidelity audio compression with improved {RVQGAN},'' in \emph{Neural Information Processing Systems (NeurIPS)}, 2023.

\bibitem{Wu2023MusicCM}
S.-L. Wu, C.~Donahue, S.~Watanabe, and N.~J. Bryan, ``{Music ControlNet}: Multiple time-varying controls for music generation,'' \emph{ArXiv}, vol. abs/2311.07069, 2023.

\bibitem{Levy2023ControllableMP}
M.~Levy, B.~D. Giorgi, F.~Weers, A.~Katharopoulos, and T.~Nickson, ``Controllable music production with diffusion models and guidance gradients,'' \emph{ArXiv}, vol. abs/2311.00613, 2023.

\bibitem{yu2023freedom}
J.~Yu, Y.~Wang, C.~Zhao, B.~Ghanem, and J.~Zhang, ``Freedom: Training-free energy-guided conditional diffusion model,'' \emph{IEEE/CVF International Conference on Computer Vision (ICCV)}, 2023.

\bibitem{Novack2024Ditto}
Z.~Novack, J.~McAuley, T.~Berg-Kirkpatrick, and N.~J. Bryan, ``{DITTO}: Diffusion inference-time t-optimization for music generation,'' 2024.

\bibitem{Wallace2023EndtoEndDL}
B.~Wallace, A.~Gokul, S.~Ermon, and N.~V. Naik, ``End-to-end diffusion latent optimization improves classifier guidance,'' \emph{IEEE/CVF International Conference on Computer Vision (ICCV)}, 2023.

\bibitem{karunratanakul2023optimizing}
K.~Karunratanakul, K.~Preechakul, E.~Aksan, T.~Beeler, S.~Suwajanakorn, and S.~Tang, ``Optimizing diffusion noise can serve as universal motion priors,'' \emph{arXiv preprint arXiv:2312.11994}, 2023.

\bibitem{song2023consistency}
Y.~Song, P.~Dhariwal, M.~Chen, and I.~Sutskever, ``Consistency models,'' \emph{arXiv preprint arXiv:2303.01469}, 2023.

\bibitem{Kim2023ConsistencyTM}
D.~Kim, C.-H. Lai, W.-H. Liao, N.~Murata, Y.~Takida, T.~Uesaka, Y.~He, Y.~Mitsufuji, and S.~Ermon, ``Consistency trajectory models: Learning probability flow {ODE} trajectory of diffusion,'' \emph{arXiv preprint arXiv:2310.02279}, 2023.

\bibitem{chen2023musicldm}
K.~Chen, Y.~Wu, H.~Liu, M.~Nezhurina, T.~Berg-Kirkpatrick, and S.~Dubnov, ``{MusicLDM}: Enhancing novelty in text-to-music generation using beat-synchronous mixup strategies,'' \emph{arXiv:2308.01546}, 2023.

\bibitem{dhariwal2021diffusion}
P.~Dhariwal and A.~Nichol, ``Diffusion models beat {GANs} on image synthesis,'' \emph{Neural Information Processing Systems (NeurIPS)}, 2021.

\bibitem{Lam2023EfficientNM}
\BIBentryALTinterwordspacing
M.~W.~Y. Lam, Q.~Tian, T.-C. Li, Z.~Yin, S.~Feng, M.~Tu, Y.~Ji, R.~Xia, M.~Ma, X.~Song, J.~Chen, Y.~Wang, and Y.~Wang, ``Efficient neural music generation,'' \emph{ArXiv}, vol. abs/2305.15719, 2023. [Online]. Available: \url{https://api.semanticscholar.org/CorpusID:258887792}
\BIBentrySTDinterwordspacing

\bibitem{Song2020ScoreBasedGM}
\BIBentryALTinterwordspacing
Y.~Song, J.~N. Sohl-Dickstein, D.~P. Kingma, A.~Kumar, S.~Ermon, and B.~Poole, ``Score-based generative modeling through stochastic differential equations,'' \emph{ArXiv}, vol. abs/2011.13456, 2020. [Online]. Available: \url{https://api.semanticscholar.org/CorpusID:227209335}
\BIBentrySTDinterwordspacing

\bibitem{Pasini2022MusikaFI}
\BIBentryALTinterwordspacing
M.~Pasini and J.~Schl{\"u}ter, ``Musika! fast infinite waveform music generation,'' in \emph{International Society for Music Information Retrieval Conference}, 2022. [Online]. Available: \url{https://api.semanticscholar.org/CorpusID:251643496}
\BIBentrySTDinterwordspacing

\bibitem{Lu2022DPMSolverFS}
C.~Lu, Y.~Zhou, F.~Bao, J.~Chen, C.~Li, and J.~Zhu, ``{DPM-Solver++}: Fast solver for guided sampling of diffusion probabilistic models,'' \emph{ArXiv}, vol. abs/2211.01095, 2022.

\bibitem{luo2023latent}
S.~Luo, Y.~Tan, L.~Huang, J.~Li, and H.~Zhao, ``Latent consistency models: Synthesizing high-resolution images with few-step inference,'' \emph{arXiv preprint arXiv:2310.04378}, 2023.

\bibitem{Sauer2023AdversarialDD}
\BIBentryALTinterwordspacing
A.~Sauer, D.~Lorenz, A.~Blattmann, and R.~Rombach, ``Adversarial diffusion distillation,'' \emph{ArXiv}, vol. abs/2311.17042, 2023. [Online]. Available: \url{https://api.semanticscholar.org/CorpusID:265466173}
\BIBentrySTDinterwordspacing

\bibitem{bai2023accelerating}
Y.~Bai, T.~Dang, D.~Tran, K.~Koishida, and S.~Sojoudi, ``Accelerating diffusion-based text-to-audio generation with consistency distillation,'' \emph{arXiv preprint arXiv:2309.10740}, 2023.

\bibitem{Si2023FreeUFL}
C.~Si, Z.~Huang, Y.~Jiang, and Z.~Liu, ``{FreeU}: Free lunch in diffusion u-net,'' \emph{ArXiv}, 2023.

\bibitem{chen2016training}
T.~Chen, B.~Xu, C.~Zhang, and C.~Guestrin, ``Training deep nets with sublinear memory cost,'' \emph{arXiv preprint arXiv:1604.06174}, 2016.

\bibitem{ho2022classifier}
J.~Ho and T.~Salimans, ``Classifier-free diffusion guidance,'' in \emph{NeurIPS Workshop on Deep Gen. Models and Downstream Applications}, 2021.

\bibitem{zheng2024trajectory}
J.~Zheng, M.~Hu, Z.~Fan, C.~Wang, C.~Ding, D.~Tao, and T.-J. Cham, ``Trajectory consistency distillation,'' \emph{arXiv preprint arXiv:2402.19159}, 2024.

\bibitem{Zhu2024MusicHiFiFH}
\BIBentryALTinterwordspacing
G.~Zhu, J.-P. Caceres, Z.~Duan, and N.~J. Bryan, ``{MusicHiFi}: Fast high-fidelity stereo vocoding,'' 2024. [Online]. Available: \url{https://api.semanticscholar.org/CorpusID:268510221}
\BIBentrySTDinterwordspacing

\bibitem{dhariwal2020jukebox}
P.~Dhariwal, H.~Jun, C.~Payne, J.~W. Kim, A.~Radford, and I.~Sutskever, ``Jukebox: A generative model for music,'' \emph{arXiv:2005.00341}, 2020.

\bibitem{wu2023large}
Y.~Wu, K.~Chen, T.~Zhang, Y.~Hui, T.~Berg-Kirkpatrick, and S.~Dubnov, ``Large-scale contrastive language-audio pretraining with feature fusion and keyword-to-caption augmentation,'' in \emph{IEEE International Conference on Audio, Speech and Signal Processing (ICASSP)}, 2023.

\bibitem{fadtk}
A.~Gui, H.~Gamper, S.~Braun, and D.~Emmanouilidou, ``Adapting {Frechet Audio Distance} for generative music evaluation,'' in \emph{IEEE International Conference on Audio, Speech and Signal Processing (ICASSP)}, 2024.

\end{thebibliography}

\end{document}